\documentclass[prc,twocolumn,nofootinbib]{revtex4}

\usepackage{amsmath,amssymb}
\usepackage{epsfig}
\usepackage{color}
\usepackage[utf8]{inputenc}
\usepackage{ulem}

\inputencoding{utf8}

\begin{document}
	
\title{Trineutron resonances in the SS-HORSE extension of the no-core shell model}

\author{I. A. Mazur}
\affiliation{Center for Exotic Nuclear Studies, Institute for Basic Science, Daejeon 34126, Republic of Korea}

\author{M. K. Efimenko}
\affiliation{Laboratory for Modeling of Quantum Processes, Pacific National University, Khabarovsk 680035, Russia}

\author{A. I. Mazur}
\affiliation{Laboratory for Modeling of Quantum Processes, Pacific National University, Khabarovsk 680035, Russia}

\author{I. J. Shin}
\affiliation{Institute for Rare Isotope Science, Institute for Basic Science, Daejeon 34000, Republic of Korea}

\author{V. A. Kulikov}
\affiliation{Skobeltsyn Institute of Nuclear Physics, Lomonosov Moscow State University, Moscow 119991, Russia}

\author{A. M. Shirokov}
\affiliation{Skobeltsyn Institute of Nuclear Physics, Lomonosov Moscow State University, Moscow 119991, Russia}

\author{J. P. Vary}
\affiliation{Department of Physics and Astronomy, Iowa State University, Ames, Iowa 50011-3160, USA}

\begin{abstract}

The SS-HORSE--NCSM method is generalized 
to the case of democratic decay into an odd number of fragments. This method is applied to the search 
for resonances in three-neutron system (trineutron) using \textit{ab initio} no-core shell model 
calculations with realistic nucleon-nucleon ($NN$) potentials. The $3/2^-$ and 
$1/2^-$ strongly overlapping resonances are predicted when 
softened $NN$ interactions are used
and are preferred over the case where
bare $NN$ interactions of
the chiral effective field theory
are used with no resonance obtained.
\end{abstract}

\maketitle

\section{Introduction}

In this paper, we develop and apply an \textit{ab initio} method of calculating the democratic decay of light nuclei  into an odd
number of fragments within the no-core shell model (NCSM)~\cite{Barrett_ProgrPartNuclPhys_2013}. 
Such an approach is of a current interest for the 
studies of neutron-excess light nuclei and, in particular, 
Borromean  neutron-excess
nuclei near and beyond the neutron drip line.

We apply this method to the search for resonant states in the three-neutron 
system (trineutron).
There is an increasing interest in theoretical and 
experimental investigations of multi-neutron systems 
following the experimental observation of the tetraneutron
resonance~\cite{Kisamori,Duer_Nature_2022}. 
By studying the multi-neutron systems we can sensitively probe the
interaction between neutrons 
for details that are not available from neutron-neutron scattering experiments.

The first experimental investigations of the three-neutron system were published
in the 1960s. In particular,
the bound trineutron search failed in the studies of the 
$^3$H($n$,$p$)$^3n$ reaction in Ref.~\cite{Ajdacic_PhysRevLett_1965}. A comprehensive description
of the history of trineutron 
experimental searches can be found in reviews
of Refs.~\cite{Kezerashvili_arXiv_2016,Marques_EurPhysJA_2021}. 
The main conclusion of all experiments is the exclusion of the bound trineutron. At the same time, 
the existence of a resonant trineutron state is not ruled out.

References~\cite{Kezerashvili_arXiv_2016,Marques_EurPhysJA_2021}
present also the history of theoretical 
investigations of the three-neutron system. Among those we note the
recent studies based on realistic $NN$ 
interactions~\cite{Hiyama_PhysRevC_2016, Gandolfi, Deltuva_PhysRevC_2018, Li_PhysRevC_2019}. 
The resonant trineutron has not 
been found in {Refs.}~\cite{Hiyama_PhysRevC_2016,Deltuva_PhysRevC_2018}. 
The binding energy of three neutrons confined by an 
external potential (trap) has been extrapolated in Ref.~\cite{Gandolfi}
to the case of the vanishing trap to estimate the trineutron resonance energy (without 
any estimation for the resonance 
width). The obtained resonance energy of $E_r=1.11(21)$~MeV
 is close to the result of Ref.~\cite{Li_PhysRevC_2019} 
where the trineutron resonance is predicted by the calculations in the
 \textit{ab initio} no-core Gamow shell model
at the energy of $E_r=1.29$~MeV with the width of $\Gamma=0.91$~MeV.

In this work we will extend our SS-HORSE--NCSM 
approach~\cite{Shirokov_PhysRevC_2016,Mazur_PhysPartNucl_2017,Blokhintsev_YaF_2017_1,Blokhintsev_YaF_2017_2,Shirokov_PhysRevC_2018,Shirokov_PhysRevLett_2016,Shirokov_AIPConfProc_2018} 
that generalizes the NCSM to the description of continuum spectrum states. The advantage of 
the SS-HORSE--NCSM is that the scattering phase shifts are computed by simple analytical expressions
at the NCSM eigenenergies and there is no need in additional numerical challenges for no-core systems   
as compared to other continuum generalizations of NCSM like the NCSM with continuum 
(NCSMC)~\cite{NCSMC} or the no-core Gamow shell model~\cite{GSM,Li_PhysRevC_2019}. 
Next the $S$ matrix is parametrized and the resonant energies and widths are
obtained by a numerical location of the $S$-matrix poles.
Recently this method has been successfully applied to
the description of resonant states in 
$^5$He~\cite{Shirokov_PhysRevC_2016}, $^5$Li~\cite{Shirokov_PhysRevC_2018}, 
$^7$He~\cite{Mazur_PhysRevC_2022}, and $^9$Li~\cite{Mazur_PhysAtNucl_2022} with 
JISP16~\cite{Shirokov_PhysLettB_2007} and Daejeon16~\cite{Shirokov_PhysLettB_2016} realistic 
$NN$ interactions in the channels of elastic scattering of protons in the case of  $^5$Li or neutrons
in all other nuclei by the remaining nuclear fragment in the ground and sometimes in  excited states. 
This method has been also generalized 
to the case of four-body democratic decays and applied to 
the description of resonances in the 
tetraneutron~\cite{Shirokov_PhysRevLett_2016, Shirokov_AIPConfProc_2018}
and in the $^7$He nucleus in the channel of four-body decay into 
$^4$He and three neutrons~\cite{Mazur_PhysRevC_2022}. In short, we have previously applied
the SS-HORSE--NCSM approach up to now only to the decay channels with
an even number of fragments.

On the other hand, the SS-HORSE--NCSM has been applied to
the hypernuclear system $\Lambda nn$ in 
Ref.~\cite{Htun_PhysRevC_2022}, a three-body decay,
for the first time. The distinction from 
Refs.~\cite{Shirokov_PhysRevC_2016,Mazur_PhysPartNucl_2017,Blokhintsev_YaF_2017_1,Blokhintsev_YaF_2017_2,Shirokov_PhysRevC_2018,Shirokov_PhysRevLett_2016,Shirokov_AIPConfProc_2018} 
is that Ref.~\cite{Htun_PhysRevC_2022} did not search for the
$S$-matrix poles but extracted the resonance parameters from the slope of the phase shifts
of the true three-body ($3\to3$) scattering. 

In this paper we generalize the technique of locating the $S$-matrix
poles proposed in
Refs.~\cite{Shirokov_PhysRevC_2016,Mazur_PhysPartNucl_2017,Blokhintsev_YaF_2017_1,Blokhintsev_YaF_2017_2,Shirokov_PhysRevC_2018,Shirokov_PhysRevLett_2016,Shirokov_AIPConfProc_2018} 
to the case of democratic decay into an odd number of
fragments. We construct a family of parametrizations of the ${3\to3}$
scattering $S$~matrix  in a minimal approximation
to enable the possibility of the $S$-matrix pole search.

The structure of the paper is the following. We discuss the $3\to3$ scattering,
the structure of the respective $S$~matrix and the generalization of the SS-HORSE--NCSM approach
to the case of the
democratic decay into an odd number of fragments using minimal approximations in Sec.~\ref{sshorse3b}. 
We apply the developed method to the search of 
resonances in the three-neutron system based on the NCSM 
calculations with various realistic $NN$ interactions 
in Sec.~\ref{3n}. The conclusions are presented in Sec.~\ref{summary}.

\section{SS-HORSE--NCSM method for democratic decay into odd number of fragments}
\label{sshorse3b}

We make use of the version of the  $J$-matrix 
formalism~\cite{Heller_PhysRevA_1974,Yamani_JMathPhys_1975} in scattering theory
utilizing the harmonic oscillator basis, which is also 
known as HORSE~\cite{Bang_AnnPhys_2000}, for the generalization of the NCSM to 
the case of the continuum spectrum. 
The essence of the HORSE formalism is the division of the 
many-body Hilbert space into a finite-dimensional 
oscillator subspace where both the potential energy of the interactions between particles and their kinetic energy are
taken into account ($P$ space) and the 
remaining infinite-dimensional subspace where
only the kinetic energy is retained and the interaction is neglected ($Q$ space).
The $P$ space  conventionally  includes all many-body states with
oscillator excitation quanta, which do not exceed some
certain number $N_{\max}$. This definition is well matched with the 
NCSM where $N_{\max}$ is used to restrict the model space.

We use a generalization of the HORSE formalism to the case of the true many-body 
($A\to A$) scattering developed in 
Ref.~\cite{Zaytsev_TeorMatFiz_1998} to describe states in the 
many-body continuum. The version of HORSE for $A\to A$ scattering utilizes the ideas of the method of
hyperspherical harmonics (HH) (see, e.\,g., Refs.~\cite{Jibuti,HHbook}), 
which was widely used 
in studies of various atomic and nuclear systems, in particular, of 
the trineutron~\cite{Dzhibuti_SovJNuclPhys264_1984,Dzhibuti_SovJNuclPhys700_1984,Jibuti_NuclPhysA_1985,Dzhibuti_SovJNuclPhys_1985,Kezerashvili_Preprint_1993}.

In the case of continuum states, the HH method is an adequate tool for the description of the
so-called democratic decays of an $A$-body system when no subgroup of the $A$ particles
has a bound state. This condition appears to be satisfied for the
trineutron or tetraneutron.
The wave function dependence on the ``democratic''  hyperradius,
\begin{equation}
\label{rho}
	\rho=\sqrt{\sum_{i=1}^{A}(\mathbf{r}_i-\mathbf{R})^2},
\end{equation}
is of a primary importance within the HH approach. Here $\mathbf{r}_i$ 
are the individual 
neutron  coordinates and $\mathbf{R}$ is the center-of-mass coordinate. The remaining degrees of freedom
are described by hyperspherical functions depending on some set of $3A-4$ angles~$\Omega_{i}$ on the 
${(3A-3)}$-dimensional sphere
coupled with neutron spins and a function describing the center-of-mass motion. Both the hyperspherical 
and the hyperradial functions are characterized by the hypermomentum~$K$ and some other quantum 
numbers~$\alpha$
distinguishing different states with the same  hypermomentum, which are of no interest for
us in this research. For the states of a definite total angular momentum~$J$ and parity,
$K=K_{\min},K_{\min}+2,\ldots$, where generally $K_{\min}\geq0$ is integer, and $K_{\min}=1$
in the case of trineutron
natural (negative) parity states with~$J=3/2$ or~1/2.

In the HH approach, the 
Schr\"odinger equation takes the form of a set of coupled
equations, which is equivalent to a set of 
equations describing a multichannel scattering 
with the same threshold in all channels. Each of the equations 
includes a centrifugal term~${\cal L}({\cal L}+1)/\rho^2$, 
where the effective orbital momentum~\cite{Zaytsev_TeorMatFiz_1998}
\begin{equation}
	{\cal L} = K+\frac{3A-6}{2}.
\label{KL}
\end{equation}

We note that the NCSM calculations performed in the $P$ space utilize
a complete set of HH with ${K\leq N_{\max}+N_{\min}}$. However,
in the $Q$ space, which is associated with 
the long-range behavior of the wave functions, the HH with $K>K_{\min}$
are suppressed by the high 
centrifugal barrier. Therefore, we utilize the democratic decay minimal approximation that implies retaining 
only one HH with $K=K_{\min}$ in the $Q$-space.
So, the wave function is 
characterized by a single phase shift $\delta$ of $A\to A$ scattering. 
This phase shift can 
be calculated using NCSM eigenenergies $E_d$ obtained with given
values of $N_{\max}$ and the NCSM oscillator basis parameter~$\hbar\omega_d$ 
within the SS-HORSE--NCSM approach as~\cite{Shirokov_PhysRevLett_2016}
\begin{equation}
	\label{sshorseeigph}
	\tan\delta(E_d)=-\frac{S_{N_{\max}+N_{\min}+2,{\cal L}}(E_d)}{C_{N_{\max}+N_{\min}+2,{\cal L}}(E_d)},
\end{equation}
where $N_{\min}$ is the minimal number of oscillator quanta allowed by the 
Pauli principle, $S_{n{\cal L}}(E)$ and $C_{n{\cal L}}(E)$ are regular and irregular solutions 
for a free motion in the HORSE formalism, which
explicit analytical expressions can be found  
in Ref.~\cite{Zaytsev_TeorMatFiz_1998}. Note, $S_{n{\cal L}}(E)$ and $C_{n{\cal L}}(E)$ 
depend on the oscillator parameter $\hbar\omega$.

The accuracy of the approximation retaining the single lowest HH in the $Q$ space 
was confirmed in the studies of three-body democratic decays in 
Refs.~\cite{Lurie_IzvRossAkadNaukSerFiz_1993,Lurie_IzvRossAkadNaukSerFiz_1997, Lurie_AnnPhys_2004,Lurie_Jmatrix_book}.
We also used the minimal approximation for the democratic decay in investigations of the
four-neutron system~\cite{Shirokov_PhysRevLett_2016, Shirokov_AIPConfProc_2018}.

The $A\to A$ $S$~matrix is related to the phase shift $\delta$,
\begin{equation}
	\label{Sdelta}
	S(k)=e^{2i\delta(E)}.
\end{equation}
To study the $S$ matrix analytical properties, it is more convenient to analyze it as a function of 
the momentum~$k$ instead of the energy~$E$, 
\begin{equation}
	E=\frac{\hbar^2k^2}{2M},
\end{equation}
where $M$ is total mass of the system.

In the case of even~$A$, ${\cal L}$ is integer, and the ${A\to A}$ $S$~matrix  analytical properties are 
similar to those of two-body scattering. In particular~\cite{Baz,Newton}, 
\begin{equation}
	\label{Smin}
	S(-k)=S^{-1}(k)
\end{equation}
and
\begin{equation}
	\label{Sconj}
	S^*(k)=\frac{1}{S(k^*)},
\end{equation}
which are crucial for the $S$-matrix parametrization. The parameterized $S$~matrix can be analytically 
continued to the complex $k$ plane for the search of its poles associated with resonant and bound states. 
This technique has been used to estimate the energy and width of
the resonant state in the tetraneutron~\cite{Shirokov_PhysRevLett_2016,Shirokov_AIPConfProc_2018}.

Analytical properties of the ${A\to A}$ $S$~matrix become more complicated in case of an odd~$A$ 
due to a half-integer value of the effective angular momentum~${\cal L}$ as follows from Eq.~\eqref{KL}. 
The $S$-matrix properties in the case of arbitrary
non-integer angular momentum are discussed in Ref.~\cite{Newton}. In this case Eq.~\eqref{Smin} is generalized to
\begin{equation}
	\label{Spiturn}
	S(ke^{i\pi})=e^{2\pi i{\cal L}}S^{-1}(k)+1-e^{2\pi i{\cal L}},
\end{equation}
which holds for any complex value of $k$. As a result, for a half-integer~${\cal L}$ we have
\begin{equation}
	\label{Spiturn3b}
	S(ke^{i\pi})=-S^{-1}(k)+2 .
\end{equation}
Note that Eq.~\eqref{Sconj} is valid for any real value of angular momentum.
We attribute properties \eqref{Sconj} and \eqref{Spiturn3b} to the ${A\to A}$ 
$S$~matrix in the case of an odd~$A$. 

The $S$~matrix has multiple sheets and its properties are complicated in the case of a non-integer 
angular momentum. The $S$~matrix can be expressed as~\cite{Alfaro}
\begin{equation}
	\label{SZ}
	S(k)=\frac{Z(k)-ik^{2{\cal L}+1}e^{i\pi(2{\cal L}+1)}}{Z(k)-ik^{2{\cal L}+1}},
\end{equation}
where $Z(k)$ has the following property:
\begin{equation}
	Z(ke^{i\pi})=Z(k).
\end{equation}

Equation~\eqref{SZ} cannot be used directly in the case of a half-integer~${\cal L}$: 
according to  Ref.~\cite{Alfaro}, in this case we 
have an uncertainty of the $0/0$ type that should be resolved using the L'H\^opital's theorem considering 
${\cal L}$ as a continuous variable and investigate the limit ${\cal L}\to K+(3A-6)/2$ to obtain
\begin{equation}
	\label{SY}
	S(k)=1+\frac{2\pi k^{2{\cal L}+1}}{Y(k)-2ik^{2{\cal L}+1}\ln(k/q_0)},
\end{equation}
where
$Y(k)=\left.\frac{\partial Z(k)}{\partial{\cal L}}\right|_{{\cal L}=K+(3A-6)/2}$ and $q_0$ is 
a real-valued momentum needed to make dimensionless the argument of~$\ln$ in the denominator. We note that our final
results for the $S$-matrix poles are independent of~$q_0$.
Using Eqs.~\eqref{Sdelta} and~\eqref{SY}, it is easy to deduce
\begin{equation}
	\label{tandeltaY}
	\tan\delta=\frac{\pi k^{2{\cal L}+1}}{2k^{2{\cal L}+1}\ln(k/q_0)+i\left(Y(k)+\pi k^{2{\cal L}+1}\right)} .
\end{equation}
The phase shift is a real-valued function for real~$k>0$. 
Therefore it is convenient to introduce a real-valued at real $k$ function
\begin{equation}
	\label{Xdef}
	X(k)=i\left(Y(k)+\pi k^{2{\cal L}+1}\right)\! .
\end{equation}
It is easy to show that $Y(ke^{i\pi})=Y(k)$, 
that leads to the following symmetry property of the function~$X(k)$:
\begin{equation}
	\label{Xeven}
	X(ke^{i\pi})=X(k).
\end{equation}
The $A\to A$ $S$~matrix and phase shift are expressed in terms of $X(k)$ as
\begin{equation}
	\label{SX}
	S(k)=\frac{X(k)+2k^{2{\cal L}+1}\ln(k/q_0)+i\pi k^{2{\cal L}+1}}{X(k)+2k^{2{\cal L}+1}\ln(k/q_0)-i\pi k^{2{\cal L}+1}},
\end{equation}
\begin{equation}
	\label{tandeltaX}
	\tan\delta=\frac{\pi k^{2{\cal L}+1}}{2k^{2{\cal L}+1}\ln(k/q_0)+X(k)}.
\end{equation}
The expression \eqref{SX} satisfies the properties of Eqs.~\eqref{Sconj} 
and~\eqref{Spiturn3b}.

Due to Eq.~\eqref{Xeven}, the function~$X(k)$ can be parameterized 
as a  series expansion in even powers of~$k$,
\begin{equation}
	\label{Xpoli}
	X(k)=\sum_{i=0}^{W}w_ik^{2i}.
\end{equation}
We note that the value of $q_0$ is arbitrary. Redefining $q_0$ results in a redefinition of
parameters $w_i$ ($i=0,\ldots,W$) in Eq.~\eqref{Xpoli} such that the $S$~matrix defined by Eq.~\eqref{SX} 
remains unchanged.

The parametrization \eqref{Xpoli} provides for an estimation of
the phase-shift behavior in the limit~$k\to0$. 
For example, for the three-body problem ($A=3$), supposing that~$X(k)\xrightarrow{k\to0}w_0$,
we obtain from Eq.~\eqref{tandeltaX}:
\begin{equation}
\tan\delta\sim\delta\sim k^{2K+4}\sim k^{2{\cal L}+1}\sim E^{K+2\!} .
\end{equation}
This behavior is in line with the analysis presented in Ref.~\cite{Jibuti} justifying the
parameterization~\eqref{Xpoli}.  

Following the ideas of the SS-HORSE--NCSM approach~\cite{Shirokov_PhysRevC_2016,Mazur_PhysPartNucl_2017,Blokhintsev_YaF_2017_1,Blokhintsev_YaF_2017_2,Shirokov_PhysRevC_2018,Shirokov_PhysRevLett_2016,Shirokov_AIPConfProc_2018}, 
we can obtain the parameters~$w_i$ of the expansion~\eqref{Xpoli} by calculating a set of the $A\to A$ phase shifts~$\delta(E_d)$ 
using Eq.~\eqref{sshorseeigph} at the
NCSM eigenenergies~$E_d$ 
obtained with a chosen~$N_{\max}$ and  a set of the~$\hbar\omega_d$ values,  and next
parameterize this set of~$\delta(E_d)$ by means of Eqs.~\eqref{tandeltaX} and~\eqref{Xpoli} (see the next section for more details).
To calculate energies and widths of resonances, we
locate the $S$-matrix poles by searching for zeros of the denominator in the right-hand side of Eq.~\eqref{SX}, which is equivalent to
solving numerically in the complex $k$ plane 
$[-\pi<\arg(k)<\pi]$ equation
\begin{equation}
	\label{SXpole}
	X(k)+2k^{2{\cal L}+1}\ln(k/q_0)-i\pi k^{2{\cal L}+1}=0
\end{equation}
using the technique suggested in Ref.~\cite{Shirokov_PhysRevC_2018} or 
the Newton--Raphson method (see, e.\,g., Ref.~\cite{Rakityansky}).

\section{Trineutron}
\label{3n}

The above method is applied to the search of resonances in the three-neutron system. 
We use various realistic $NN$ interactions, the same as employed in our 
analysis of the tetraneutron~\cite{Shirokov_PhysRevLett_2016, Shirokov_AIPConfProc_2018}. 
We utilize the MFDn code~\cite{Maris_ProcComputSci_2010, Aktulga_ConcurComputPractExper_2014} 
to perform the NCSM calculations with $N_{\max}$ ranging from 4--20 and $\hbar\omega$ spanning from 2--50~MeV.

\begin{figure}[t]
	\epsfig{file=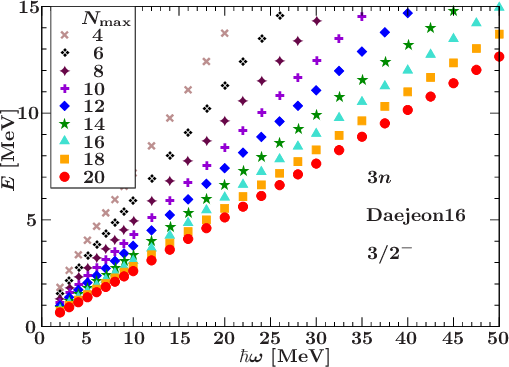,width=0.5\textwidth}
	\epsfig{file=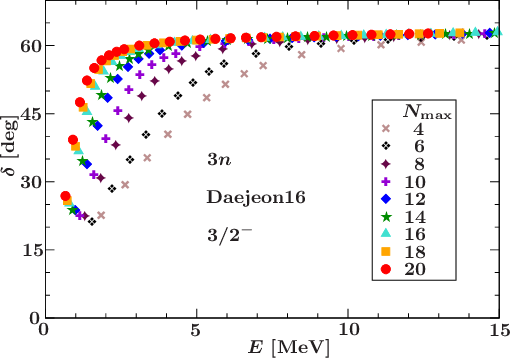,width=0.5\textwidth}
	\caption{Top: NCSM results for the trineutron $3/2^{-}$ ground state energy obtained with Daejeon16 
		$NN$ interaction with various $N_{\max}$ plotted as functions of $\hbar\omega$. Bottom: $3\to3$ 
		phase shifts at the NCSM eigenenergies obtained using Eq.~\eqref{sshorseeigph}.}
	\label{Daej_all}
\end{figure}

The results for the $3/2^-$ ground-state energy obtained with the Daejeon16~\cite{Shirokov_PhysLettB_2016} interaction, are shown
in the top panel of Fig.~\ref{Daej_all}. The $3\to3$ phase shifts at the NCSM eigenenergies obtained 
using Eq.~\eqref{sshorseeigph} are presented in the bottom panel. It is seen that the phase shifts tend to the same smooth 
resonance-like curve as $N_{\max}$ is increasing demonstrating a convergence of the $3\to3$ phase shift calculations.

We parametrize the function~$X(k)$ for each individual value of~${N_{\max}\ge16}$ used in the NCSM
calculations of the trineutron ground-state energies. For a given~$N_{\max}$,
we use a set of parameters~$w_{i}$ defining~$X(k)$ to find the energies~${\cal E}_d$ by solving the equation 
\begin{equation}
	\label{ENhw_implicit}
	-\frac{S_{N_{\max}+3,{\cal L}}({\cal E}_d)}{C_{N_{\max}+3,{\cal L}}({\cal E}_d)}=\frac{\pi\kappa^6_d}{2\kappa^6_d\ln(\kappa_d/q_0)+X(\kappa_d)},
\end{equation}
for each value of~$\hbar\omega_{d}$ used in the respective NCSM calculations. Here
$\kappa_d=\frac{\sqrt{2M{\cal E}_d}}{\hbar}$ and ${\cal L}=K_{\min}+\frac32=\frac52$. 
To find the optimal values of $w_i$, we minimize the function
\begin{equation}
	\label{Xi}
	\Xi=\sqrt{\frac1D\sum_{d=1}^D\left[({\cal E}_d-E_d)^2\left(\frac{\hbar\omega_M}{\hbar\omega_d}\right)^2\right]},
\end{equation}
where $D$ is the number of the NCSM energies $E_d$ obtained with the same $N_{\max}$ and the same $\hbar\omega_d$
as each of the respective energies~${\cal E}_d$, $\displaystyle\hbar\omega_M=\max_{d=1,\ldots,D}\hbar\omega_d$, and 
$\left(\hbar\omega_M/\hbar\omega_d\right)$ is the weight increasing the importance of states with smaller~$\hbar\omega_d$
corresponding to energies closer to the resonance region.

The quality of the fits can be estimated by the r.m.s. deviation
\begin{equation}
	\label{rms}
	\xi=\sqrt{\frac1D\sum_{d=1}^D({\cal E}_d-E_d)^2}.
\end{equation}
In our case we get approximately the same r.m.s.\ deviations~$\xi$ obtained 
with five ($W=4$) or six ($W=5$) parameters~$w_{i}$ in Eq.~\eqref{Xpoli}, which are, however, significantly smaller than the r.m.s.\ deviations
obtained with four parameters ($W=3$).
Energies and widths obtained by locating the $S$-matrix poles using Eq.~\eqref{SXpole} and the NCSM results from calculations
with $N_{\max}=16$, 18, 20 and parametrizations with~$W=4$ and~5 together with the respective~$\xi$ values are presented in
Table~\ref{Tab:convergence32mDaej}.

\begin{table}[t]
	\caption{Convergence of	energy $E_r$ and width $\Gamma$ of the trineutron $3/2^-$ resonant state 
		obtained with $NN$ interaction Daejeon16 with increasing $N_{\max}$. $\xi$ is the r.m.s. 
		deviation defined by Eq.~\eqref{rms}.}
	\begin{ruledtabular}
		\begin{tabular}{c|ccc|ccc}
			$N_{\max}$   & 16    & 18    & 20    & 16    & 18    & 20    \\
			$W$          &  4    &  4    &  4    &  5    & 5     &  5    \\
			\hline
			$E_r$, MeV   & 0.560 & 0.508 & 0.483 & 0.607 & 0.537 & 0.481 \\
			$\Gamma$, MeV& 1.458 & 1.152 & 0.924 & 1.524 & 1.176 & 0.963 \\
			$\xi$, keV   & 3.3   & 3.9   & 2.5   & 2.7   & 2.0   & 1.8   \\
		\end{tabular}
	\end{ruledtabular}
	\label{Tab:convergence32mDaej}
\end{table}

\begin{figure}[t]
	\epsfig{file=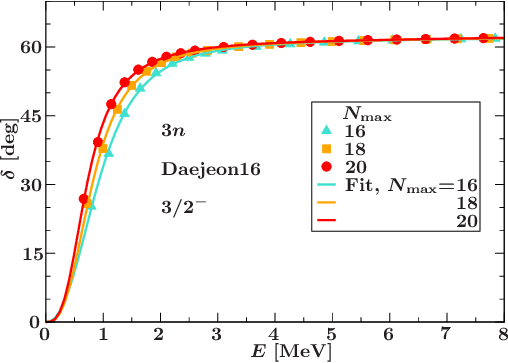,width=0.5\textwidth}
	\caption{Fits of $3\to 3$ phase shifts in the $3/2^-$ trineutron state obtained with $NN$ interaction 
		Daejeon16 and~${W=5}$ in the $X(k)$ expansion~\eqref{Xpoli}.}
	\label{Daej_s2_50}
\end{figure}

Fits of the $3\to3$ phase shifts in the $3/2^{-}$ state with six~parameters~$w_{i}$ in Eq.~\eqref{Xpoli} ($W=5$) are presented
by solid curves in Fig.~\ref{Daej_s2_50}.
The $3\to3$ phase shifts at the NCSM eigenenergies obtained 
by Eq.~\eqref{sshorseeigph} and used for the 
fitting are shown by symbols in Fig.~\ref{Daej_s2_50}.
The trineutron resonance energy and width obtained by locating the $S$-matrix pole based on the NCSM calculation
with $N_{\max} = 20$ and fit with~${W = 5}$ are adopted as the final result
presented in the 
Table~\ref{Tab:OurResult} together with their
uncertainties estimated as 
deviations of results obtained with $N_{\max}=18$, 20 and $W=4$, 5 
from the final result.

It is interesting that  we obtain in the trineutron NCSM calculations  the $1/2^-$ state very close to 
the lowest $3/2^-$ state. We perform the same analysis of the $1/2^-$ trineutron resonance. The $3\to3$ 
$1/2^-$ phase shifts are very close to those in the $3/2^-$ state and  the obtained $1/2^-$ resonance energy 
and width are presented in Table~\ref{Tab:OurResult}. It is seen that the $3/2^-$ and $1/2^-$ resonance energies 
and widths are the same within the uncertainty estimations and these resonances completely overlap.

We employ the same technique to search for resonances with other soft $NN$ interactions, in particular, with 
Idaho N$^3$LO~\cite{Entem_PhysRevC_2003} softened by the similarity renormalization group (SRG) 
transformation~\cite{Glazek_PhysRevD_1993, Wegner_AnnPhys_1994} with the flow parameter 
$\Lambda=2$~fm$^{-1}$ and JISP16~\cite{Shirokov_PhysLettB_2007}. The respective results are also presented 
in the Table~\ref{Tab:OurResult}. Note, in these cases the $3/2^-$ and $1/2^-$ resonance also degenerate 
and strongly overlap.

\begin{table}[b]
	\caption{Energies $E_r$ and widths $\Gamma$ of trineutron resonant states obtained with soft $NN$ interactions 
		Daejeon16~\cite{Shirokov_PhysLettB_2016}, JISP16~\cite{Shirokov_PhysLettB_2007}, and SRG-evolved 
		Idaho N$^3$LO~\cite{Entem_PhysRevC_2003}. 
		Uncertainties are presented in parentheses. All values are in MeV.}
	\begin{ruledtabular}
		\begin{tabular}{c|cc|cc}
			&\multicolumn{2}{c|}{$3/2^-$}&\multicolumn{2}{c}{$1/2^-$}\\
			Interaction	 & $E_r$   & $\Gamma$ & $E_r$    & $\Gamma$  \\
			\hline
			Daejeon16    & 0.48(6) & 0.96(21) & 0.48(8)  & 0.96(17)  \\
			JISP16       & 0.35(8) & 0.70(9) & 0.35(11) & 0.67(22)  \\
			\parbox{1.8cm}{N$^3$LO, SRG,	$\Lambda=2$ fm$^{-1}$}&0.34(8)&0.70(19)& 0.35(9)& 0.68(16)\\
		\end{tabular}
	\end{ruledtabular}
	\label{Tab:OurResult}
\end{table}

We also have analyzed the trineutron resonance with bare realistic $NN$ interactions derived in 
chiral effective field theory: Idaho N$^3$LO and LENPIC N$^4$LO semi-local coordinate space 
interaction~\cite{Epelbaum_PhysRevLett_2015} with regulator $R=0.9$~fm. In these cases, 
the $3\to3$ phase shifts do not demonstrate a resonant behavior.
Due to the almost complete degeneracy 
of the $3/2^-$ and $1/2^-$ states, these interactions also do not support the $1/2^-$ resonance.

We obtained~\cite{Shirokov_AIPConfProc_2018} a $4\to4$ $S$-matrix pole at negative imaginary momentum
in the tetraneutron calculations with the Idaho N$^3$LO interaction, which corresponds to a virtual state at the
energy of $E_v=-15.2$~keV. It is easy to prove, that the $A\to A$ 
$S$-matrix for an odd number of fragments~$A$ does not allow $S$-matrix poles at 
the negative imaginary half-axis of momentum. Therefore, a virtual state is prohibited in 
the trineutron treated as a democratic decaying system.

\section{Summary and conclusions}
\label{summary}

We suggest an extension of the SS-HORSE--NCSM method to a democratic decay into odd number 
of fragments. The first application of this method is the analysis of the resonant trineutron state.

We conclude that the soft $NN$ interactions that we investigated predict
two low-lying  nearly degenerate overlapping trineutron resonances with spin-parities~$3/2^-$ and $1/2^-$. 
On the other hand, these resonances are not supported by bare $NN$ interactions 
of chiral effective field theory. We do not include $NNN$ interaction in our calculations, which has
yet to be designed for three-nucleon systems with isospin $T=3/2$.

We argue that the Daejeon16 $NN$ interaction is preferable for the trineutron studies since it originates 
from the chiral effective field theory and is fitted to stable light nuclei up to 
$^{16}$O by phase-equivalent off-shell variations, which effectively mimic effects of $NNN$ forces. 
The JISP16 $NN$ interaction, which was also fitted to light nuclei by off-shell variations, leads 
to similar trineutron results as well as the SRG-evolved Idaho N$^3$LO $NN$ interaction.

We predict two overlapping trineutron resonances with spin-parities $3/2^-$ and $1/2^-$ with  nearly exactly 
the same energies~$E_{r}$
and widths~$\Gamma$: $E_{r}\simeq 0.5$~MeV and $\Gamma\simeq 1$~MeV obtained in calculations with the
Daejeon16 
and ${E_{r}\simeq 0.35}$~MeV and $\Gamma\simeq 0.7$~MeV obtained in calculations with the
JISP16 and SRG-evolved N$^{3}$LO $NN$ interactions.

Our results are in line with the conclusions of Refs.~\cite{Gandolfi,Li_PhysRevC_2019} predicting 
the trineutron resonance at lower energy 
than the tetraneutron resonance~\cite{Shirokov_PhysRevLett_2016,Shirokov_AIPConfProc_2018}.
However, in our calculations we obtain the trineutron resonance at lower energies as 
compared to $E_r=1.29$~MeV in Ref.~\cite{Li_PhysRevC_2019} and $E_r=1.11(21)$~MeV in Ref.~\cite{Gandolfi}. 
Meanwhile, its width obtained with the Daejeon16 $NN$ interaction is in agreement with $\Gamma=0.91$~MeV 
proposed in Ref.~\cite{Li_PhysRevC_2019}.
We note that Refs.~\cite{Gandolfi,Li_PhysRevC_2019} do not specify the spin-parity of the predicted 
trineutron resonance. 

\textit{Note added in proof}. Recently, a new unsuccessful experimental attempt to find the trineutron 
and triproton resonances in the reactions $^3$H($t$,$^3$He)$3n$ and $^3$He($^3$He,$t$)$3p$ was 
published in Ref.~\cite{Miki_PhysRevLett_2024}.

\section*{Acknowledgments}

The work of I.~A.~M. is supported by the Institute for Basic Science (IBS-R031-D1).
The work of M.~K.~E. and A.~I.~M. is supported by the Ministry of Science and Higher Education of Russian Federation (project No.~FEME-2024-0005).
The work of I.~J.~S. is supported by the National Research Foundation of Korea (NRF) funded by Ministry of Science and ICT (2013M7A1A1075764).
A.~M.~S. is thankful to the Chinese Academy of Sciences President's International Fellowship Initiative Program (Grant No.~2023VMA0013)
which supported his visits to the Institute of Modern Physics, Chinese Academy of Sciences in Lanzhou, China, where a part of this work was performed 
and acknowledges the hospitality of Chinese colleagues during these visits.
The work of J.~P.~V. is supported by the U.S. Department of Energy, Division of Nuclear Physics, Grant No. DE-SC0023692.


\begin{thebibliography}{99}
	
\bibitem{Barrett_ProgrPartNuclPhys_2013} B. R. Barrett, P. Navr\'atil, and J. P. Vary, Prog. Part. Nucl. Phys. 69, 131 (2013).
\bibitem{Kisamori} K. Kisamori {\it et al.}, Phys. Rev. Lett. {\bf 116}, 052501 (2016).

\bibitem{Duer_Nature_2022} M. Duer \textit{et al.}, Nature \textbf{606}, p. 678 (2022).

\bibitem{Ajdacic_PhysRevLett_1965} V. Ajda\v{c}i\'c, M. Cerineo, B. Lalovi\'c, G. Pai\'c, I. \v{S}laus, and P. Toma\v{s}, Phys. Rev. Lett. \textbf{14}, 444 (1965).
\bibitem{Kezerashvili_arXiv_2016} R. Kezerashvili, in \textit{Fission and Properties of Neutron-Rich Nuclei} (World Scientific, Singapore, 2017), p. 403.
\bibitem{Marques_EurPhysJA_2021} F. M. Marqu\'es and J. Carbonell, Eur. Phys. J. A \textbf{57}, 105 (2021).
	
\bibitem{Hiyama_PhysRevC_2016} E. Hiyama, R. Lazauskas, J. Carbonell, M. Kamimura, Phys. Rev. C \textbf{93}, 044004 (2016).
\bibitem{Gandolfi} S. Gandolfi, H.-W. Hammer, P. Klos, J. E. Lynn, and A.~Schwenk, Phys. Rev. Lett. \textbf{118}, 232501 (2017).
\bibitem{Deltuva_PhysRevC_2018} A. Deltuva, Phys. Rev. C \textbf{97}, 034001 (2018).
\bibitem{Li_PhysRevC_2019} J. G. Li, N. Michel, B. S. Hu, W. Zuo, and F. R. Xu, Phys. Rev. C \textbf{100}, 054313 (2019).
	
\bibitem{Shirokov_PhysRevC_2016} A. M. Shirokov, A. I. Mazur, I. A. Mazur, and J. P. Vary, Phys. Rev. C \textbf{94}, 064320 (2016).
\bibitem{Mazur_PhysPartNucl_2017} I. A. Mazur, A. M. Shirokov, A. I. Mazur, and J. P. Vary, Phys. Part. Nucl. \textbf{48}, 84 (2017).
\bibitem{Blokhintsev_YaF_2017_1} L. D. Blokhintsev, A. I. Mazur, I. A. Mazur, D. A. Savin, and A. M. Shirokov, Yad. Fiz. \textbf{80}, 102 (2017); Phys. Atom. Nucl. \textbf{80}, 226 (2017).
\bibitem{Blokhintsev_YaF_2017_2} L. D. Blokhintsev, A. I. Mazur, I. A. Mazur, D. A. Savin, and A. M. Shirokov, Yad. Fiz. \textbf{80}, 619 (2017); Phys. Atom. Nucl. \textbf{80}, 1093 (2017).
\bibitem{Shirokov_PhysRevC_2018} A. M. Shirokov, A. I. Mazur, I. A. Mazur, E. A. Mazur, I. J. Shin, Y. Kim, L. D. Blokhintsev, and J. P. Vary, Phys. Rev. C \textbf{98}, 044624 (2018).
\bibitem{Shirokov_PhysRevLett_2016} A. M. Shirokov, G. Papadimitriou, A. I. Mazur, I.~A.~Mazur, R. Roth, and J. P. Vary, Phys. Rev. Lett. \textbf{117}, 182502 (2016).
\bibitem{Shirokov_AIPConfProc_2018} A. M. Shirokov, Y. Kim, A. I. Mazur, I. A. Mazur, I.~J.~Shin, and J. P. Vary, AIP Conf. Proc. \textbf{2038}, 020038 (2018).
\bibitem{NCSMC}P. Navr\'atil, S. Quaglioni, G. Hupin, C. Romero-Redondo, and A. Calci, Phys. Scr. {\bf 91}, 053002 (2016).
\bibitem{GSM}G. Papadimitriou, J. Rotureau, N. Michel, M.~Płoszajczak, and B. R. Barrett, Phys. Rev. C {\bf 88}, 044318 (2013).

\bibitem{Mazur_PhysRevC_2022} I. A. Mazur, I. J. Shin, Y. Kim, A. I. Mazur, A.~M.~Shirokov, P. Maris, and J. P. Vary, Phys. Rev. C \textbf{106}, 064320 (2022).
\bibitem{Mazur_PhysAtNucl_2022} I. A. Mazur, A. I. Mazur, V. A. Kulikov, A. M. Shirokov, I. J. Shin, Y. Kim, P. Maris, and J. P. Vary, Yad. Fiz. \textbf{86}, 104 (2023); Phys. At. Nucl. \textbf{85}, 823 (2022).
	
\bibitem{Shirokov_PhysLettB_2007} A. M. Shirokov, J. P. Vary, A. I. Mazur, and T.~A.~Weber, Phys. Lett. B \textbf{644}, 33 (2007); a Fortran code generating the JISP16 matrix elements is available at 
\url{http://lib.dr.iastate.edu/energy\_datasets/2/}.
\bibitem{Shirokov_PhysLettB_2016} A. M. Shirokov, I. J. Shin, Y. Kim, M. Sosonkina, P. Maris, and J. P. Vary, Phys. Lett. B \textbf{761}, 87 (2016); a Fortran code generating the Daejeon16 matrix elements is available at \url{http://lib.dr.iastate.edu/energy\_datasets/1/}.
	
\bibitem{Htun_PhysRevC_2022} T. Y. Htun and Y. Yan, Phys. Rev. C \textbf{105}, 064001 (2022).

\bibitem{Heller_PhysRevA_1974} E. J. Heller and H. A. Yamani, Phys. Rev. A \textbf{9}, 1201 (1974).
\bibitem{Yamani_JMathPhys_1975} H. A. Yamani and L. Fishman, J. Math. Phys. \textbf{16}, 410 (1975).
\bibitem{Bang_AnnPhys_2000} J.~M.~Bang, A.~I.~Mazur, A.~M.~Shirokov, Yu.~F.~Smirnov, and S.~A.~Zaytsev, Ann. Phys. (NY) {\bf 280}, 299 (2000).

\bibitem{Zaytsev_TeorMatFiz_1998} S. A. Zaitsev, Yu. F. Smirnov, and A. M. Shirokov, Teor. Mat. Fiz. \textbf{117}, 227 (1998); Theor. Math. Phys. \textbf{117}, 1291 (1998).
\bibitem{Jibuti} R. I. Jibuti and N. B. Krupennikova, \textit{The Method of Hyperspherical Functions in the Quantum Mechanics of Few Bodies} [in Russian] (Metsniereba, Tbilisi, 1984).
\bibitem{HHbook}J. E. Avery and J. S. Avery, {\it Hyperspherical Harmonics and Their Physical Applications}
(World Scientific, Singapore, 2018).

\bibitem{Dzhibuti_SovJNuclPhys264_1984} R. I. Dzhibuti and R. Ya. Kezerashvili, Sov. J. Nucl. Phys. \textbf{39}, 264 (1984).
\bibitem{Dzhibuti_SovJNuclPhys700_1984} R. I. Dzhibuti and R. Ya. Kezerashvili, Sov. J. Nucl. Phys. \textbf{39}, 700 (1984).
\bibitem{Jibuti_NuclPhysA_1985} R. I. Jibuti and R. Ya. Kezerashvili, Nucl. Phys. \textbf{A437}, 687 (1985).
\bibitem{Dzhibuti_SovJNuclPhys_1985} R. I. Dzhibuti and R. Ya. Kezerashvili, Sov. J. Nucl. Phys. \textbf{40}, 443 (1985).
\bibitem{Kezerashvili_Preprint_1993} R. Ya. Kezerashvili, preprint IFUP-TH 24/93, 1993.
\bibitem{Lurie_IzvRossAkadNaukSerFiz_1993} Yu. A. Lurie, Yu. F. Smirnov, and A. M. Shirokov, Izv. Ross. Akad. Nauk, Ser. Fiz. \textbf{57}, 193 (1993) [Bull. Russ. Acad. Sci., Phys. Ser. \textbf{57}, 943 (1993)].
\bibitem{Lurie_IzvRossAkadNaukSerFiz_1997} Yu. A. Lurie and A. M. Shirokov, Izv. Ross. Akad. Nauk, Ser. Fiz. \textbf{61}, 2121 (1997) [Bull. Russ. Acad. Sci., Phys. Ser. \textbf{61}, 1665 (1997)].
\bibitem{Lurie_AnnPhys_2004} Yu. A. Lurie and A. M. Shirokov, Ann. Phys. (NY) \textbf{312}, 284 (2004).
\bibitem{Lurie_Jmatrix_book} Yu. A. Lurie and A. M. Shirokov, in 
\textit{The J-Matrix Method. Developments and Applications}, edited by A. D. Alhaidari, H. A. Yamani, E. J. Heller, and M. S. Abdelmonem (Springer, New York, 2008), p. 183.
	
\bibitem{Baz} A. I. Baz’, Ya. B. Zel’dovich, and A. M. Perelomov, \textit{Scattering, Reactions and Decay in Non-relativistic Quantum Mechanics} (Israel Program for Scientific Translation, Jerusalem, 1969).
\bibitem{Newton} {R. G. Newton, {\it Scattering Theory of Waves and Particles}, 2nded. (Springer-Verlag, New York, 1982)}.
\bibitem{Alfaro} V. de Alfaro and T. Regge, \textit{Potential Scattering} (North-Holland, Amsterdam, 1965).
\bibitem{Rakityansky} S. A. Rakityansky, \textit{Jost Functions in Quantum Mechanics. A Unified Approach to Scattering, Bound, and Resonant State Problems} (Springer Nature, Cham, 2022).
	
\bibitem{Maris_ProcComputSci_2010} P. Maris, M. Sosonkina, J. P. Vary, E. G. Ng, and C.~Yang, Proc. Comput. Sci. \textbf{1}, 97 (2010).
\bibitem{Aktulga_ConcurComputPractExper_2014} H. M. Aktulga, C. Yang, E. G. Ng, P. Maris, and J.~P.~Vary, Concur. Comput. Pract. Exper. \textbf{26}, 2631 (2014).
	
\bibitem{Entem_PhysRevC_2003} D.R. Entem, R. Machleidt, Phys. Rev. C \textbf{68} (2003) 041001(R).
\bibitem{Glazek_PhysRevD_1993} S. D. Glazek and K. G. Wilson, Phys. Rev. D \textbf{48}, 5863 (1993).
\bibitem{Wegner_AnnPhys_1994} F. Wegner, Ann. Phys. (NY) \textbf{506}, 77 (1994).
\bibitem{Epelbaum_PhysRevLett_2015}  E. Epelbaum, H. Krebs, and U.-G. Mei{\ss}ner Phys. Rev. Lett. \textbf{115}, 122301 (2015).

\bibitem{Miki_PhysRevLett_2024} K. Miki \textit{et al.} (RIBF-SHARAQ11 Collaboration and RCNP-E502 Collaboration), Phys. Rev. Lett. \textbf{133}, 012501 (2024).
	
\end{thebibliography}
\end{document}